\documentclass[preprint]{aastex} 
\newcommand\etal {et~al.}
\newcommand\eg {e.g.}
\newcommand\ie {i.e.}
\newcommand\vs {{\it vs~}}
\begin{document}

\title{VLBA OBSERVATIONS OF A SAMPLE OF NEARBY FR~I RADIO GALAXIES}
\author{Chun Xu\altaffilmark{1}, Stefi A. Baum, Christopher P. O'Dea}
\affil{Space Telescope Science Institute,
       3700 San Martin Drive, Baltimore, MD 21218}
\author{J.M. Wrobel}
\affil{National Radio Astronomy Observatory\altaffilmark{2},
       P.O. Box O, Socorro, New Mexico 87801}
\author{J.J. Condon}
\affil{National Radio Astronomy Observatory\altaffilmark{2},
       520 Edgemont Road, Charlottesville, Virginia 22903}
\affil{chunxu@stsci.edu, sbaum@stsci.edu, odea@stsci.edu, 
       jwrobel@nrao.edu, jcondon@nrao.edu}
\altaffiltext{1}{Also at: Department of Astronomy, University of
Maryland, College Park, MD 20742}
\altaffiltext{2}{NRAO is a facility of the National Science Foundation
operated under cooperative agreement by Associated Universities, Inc.}

\begin{abstract}
We observed 17 nearby low luminosity FR~I radio galaxies using the NRAO
Very Long Baseline Array (VLBA) at 1.67 GHz, as part of a
multi-wavelength study of a complete sample of 21 sources selected by
radio flux density from the Uppsala General Catalogue of Galaxies.  We
detected radio emission from all 17 galaxies.  At a FWHM resolution of
$\approx 10\times 4$ mas, five galaxies show only an unresolved radio
core, ten galaxies show core-jet structures, and two galaxies show
twin-jet structures.  Comparing these VLBA images with images
previously obtained with the NRAO VLA, we find that all detected VLBA
jets are well-aligned on parsec scales with the VLA jets on kilo parsec
scales, and that the jet to counter-jet surface brightness ratios, or
the {\em sidedness\/}, decreases systematically with increasing
distance along the jet.  We attribute the sidedness to the Doppler
boosting effect and its decline to the deceleration of the jets. 
We show that a distribution of Lorentz factor centered near $\Gamma =
5$ can reproduce our VLBA detection statistics for core, core-jet and
twin-jet sources.  We also note that the luminosity per unit length,
$L_j$, of the VLBA jets drops quickly with distance, $r$, along the jet,
approximately as $L_j \propto r^{-2.0}$.  We discuss three different
mechanisms to explain this jet {\em fading\/}: (1) the decrease of
Doppler boosting due to jet deceleration, (2) synchrotron losses, and
(3) expansion losses in constant velocity but adiabatically spreading 
jets.  Mechanisms (1) and (2) are inconsistent with the observations, 
while mechanism (3) is consistent with the observations provided the 
magnetic field lines in the jets are aligned perpendicular to the jet 
axis.  This implies that the deceleration of the jets required by
the Unified Scheme does not  occur on the tens of parsec scales, but 
must occur on larger scales. 

\end{abstract}


\section{Introduction}

It is now generally agreed that there exist two separate categories of
Active Galactic Nuclei (AGN), namely radio-loud and radio-quiet.  The
former have strong radio emission and usually show extended radio
structures up to Mpc scales.  In contrast, the latter have only weak
radio emission, being about three orders of magnitude weaker at a
fixed bolometric luminosity; and tend to exhibit more compact or
slightly resolved radio structures, extending only $\le$ 1 kpc.
Within each category there are also sub-categories.  The radio-louds
can be sub-classified as FR~I and FR~II (Fanaroff \& Riley 1974; Bridle
1984).  The FR~Is show edge-darkened radio structures, while the FR~IIs
show edge-brightened radio structures.  Fanaroff \& Riley (1974) found
that radio sources with total radio powers of less than $10^{25}$
W~Hz$^{-1}$ at 408 MHz show almost exclusively FR~I morphologies, while
those with radio powers higher than $10^{27}$ W~Hz$^{-1}$ show almost
exclusively FR~II morphologies.  For sources with intermediate radio
powers, FR~I and FR~II radio morphologies are found in roughly equal
numbers.  Some ``intermediate'' radio morphologies between FR~I and
FR~II are also found to exist at those intermediate radio powers (e.g.,
Baum \etal\ 1988; Owen \& Laing 1989).  Zirbel \& Baum (1995) also
found that the difference between the core radio powers of an FR~I and
an FR~II is roughly ten times less than the difference between their
total radio powers, \ie, that FR~Is have cores which are a higher
fraction of their total luminosity than FR~IIs.

Given the complexity of AGNs, there have been many attempts to contain
the diversity under unified pictures.  For example, one of the schemes
suggests that FR~I and BL~Lacertae objects are intrinsically the same
and the only difference between them is the viewing angle (\eg, Browne
1983; Antonucci \& Ulvestad 1985; Urry \& Padovani 1995).  It has also
been suggested that FR~Is and FR~IIs have the same central engines and
their differences are induced by their different environments (\eg,
De Young 1993; Giovannini \etal\ 1994; Bicknell 1995; cf: Baum \etal\ 1995).  
The central engine in all
AGNs is believed to be a super-massive black hole which accretes mass
from its environment, as reviewed by Rees (1984).  To disentangle the
complexity of AGNs and to test different unified schemes for AGNs,
systematic multi-wavelength observations of a well defined set of AGNs
are necessary.  We have started a series of multi-wavelength
observations of a complete radio-flux-density-limited set of nearby
FR~I radio galaxies.  The instruments we employed to date include the
NRAO VLBA, the NRAO VLA, HST, and ROSAT.

Our sample consists of 21 galaxies drawn from a catalog of radio-loud
galaxies constructed by Condon \& Broderick (1988, hereafter CB88).
That catalog was based on position coincidence of radio identifications
in the Green Bank 1400 MHz sky images (Condon \& Broderick 1985, 1986;
CB88) and galaxies in the Uppsala General Catalogue of Galaxies
(Nilson 1973, hereafter referred to as UGC).  The 21 galaxies selected
for our study satisfy the following criteria: (1) Hubble type E or S0,
(2) recession velocity $v_{rec} <$ 7000 km~s$^{-1}$
(corresponds to 93 Mpc for H$_0$ = 75 km~s$^{-1}$~Mpc$^{-1}$), 
(3) optical major
axis diameter $>$ 1\arcmin, (4) total flux density $S$ at 1400 MHz $>$
150 mJy, (5) declination -5\arcdeg$\le \delta \le$ 82\arcdeg, (6)
having a ``monster'' rather than ``starburst'' energy source as
discerned from the IRAS/radio flux ratio (CB88), and (7) a radio size
$\ge$ 10\arcsec~ as measured from VLA images at 1.49 GHz with
2\arcsec~ resolution (Wrobel, Condon, \& Machalski, in preparation),
where the size is measured at contours of three times the image noise
levels.  The complete sample of 21 galaxies is listed in Table~1.  In
this paper we present the results from the VLBA observations.  The HST
images are presented by Verdoes \etal\ (1999).  Throughout this paper
we adopt H$_0$ = 75 km~s$^{-1}$~Mpc$^{-1}$ and define a spectral index
$\alpha$ as $S_{\nu}\propto {\nu}^{-\alpha}$.

We wish to emphasize that a very important feature of our project is
the proximity of the sample, so useful resolutions on AGN spatial
features can be achieved.  The farthest source in our sample is less
than 100 Mpc distant, while the closest source is about 20 Mpc away
(Table~1).  Therefore with the 100-milliarcsecond (mas) angular
resolution of HST, we resolve scales as small as $\approx$ 6 pc near the
AGN core.  The environment on these scales is directly related to the
nuclear activity.  For example, the radius at which Bondi accretion is
initiated (Bondi 1952) is $G M m_H/kT \approx 530 $ pc for a $10^8
M_{\odot}$ black hole surrounded by hot gas with temperature $10^5$ K.
Physically, Bondi accretion can be considered as the starting point
for spherically symmetric accretion, which ultimately leads to the
formation of an accretion disk.  Another important AGN feature is the
so-called Broad Line Region (BLR) with characteristic size about 1 pc
which, depending on the source, corresponds to 10 mas or $\sim
{{1}\over{10}}$ of the HST resolution.  The VLBA with an angular
resolution of $\approx$ 5 mas can image scales well inside a 1-pc BLR.

\section{VLBA Observations and Data Analysis}\label{secdata}

Fifteen galaxies from Table~1 were observed with the VLBA (Napier \etal\ 1994), 
on 1997 April 9, and two others (UGC\,00408 and UGC\,01004) were observed on
1997 August 22.  We did not observe four of the sample galaxies with the
VLBA, because UGC\,06635 lacks a VLA core, making it an unsuitable
target for the VLBA; UGC\,07654 (NGC\,4486 = 3C\,274 = M87) has been
extensively studied by others (\eg, Junor \& Biretta 1995); and
UGC\,07115 and UGC\,12064 were added into our sample after the VLBA
observations were proposed and completed.

The VLBA received data in dual circular polarizations.  Data from four
contiguous 8-MHz bands were acquired with 4-level sampling for each
polarization, providing a bandwidth of 32 MHz with a center frequency
of 1667.49 MHz.  A coordinate equinox of 2000 was assumed.
Each galaxy was observed in a phase-referencing mode (Beasley \&
Conway 1995), meaning that a galaxy observation of duration 2.5-3.0
minutes was preceded and followed by a 2-minute observation of a phase
calibrator, identified in Table~2, and typically located about
2\arcdeg~ from the galaxy.  That table also lists the assumed positions
and, where available, the 2-D positional accuracies of the phase
calibrators (Patnaik \etal\ 1992; Browne \etal\ 1998; Peck \& Beasley
1998; Wilkinson \etal\ 1998).  Observations of each galaxy were spread
in time to enhance coverage in the (u,v) plane.  On the first
observing day, each galaxy was observed 8-14 times, yielding on-source
integration times of 20-35 minutes.  On the second observing day, each
galaxy was observed 24 times, yielding on-source integration times of
72 minutes.  On both days, data were also acquired on J0555+3948
(DA\,193) and used to align the phases of the four contiguous 8-MHz
bands.  System temperatures and gains were used to set the amplitude
scale to an accuracy of about 5\%, after first applying {\em
a priori\/} editing and correcting for sampler errors.  The data on
J0555+3948 were also used to verify the amplitude calibration.  Only
parallel-hand correlations were delivered and archived at the VLBA
correlator.  All editing, amplitude calibration, fringe-fitting, phase
calibration, and imaging was done with the NRAO AIPS software.

The positions given in Table~1 for UGC\,00597 (NGC\,315) and
UGC\,07494 (NGC\,4374 = 3C\,272.1 = M84) are from M.\ Eubanks (private
communication), while the position for UGC\,07654 is from Ma \etal\
(1998); these positions carry 2-D errors of $\le$ 2 mas.  For the
other 15 galaxies observed with the VLBA, phase-referenced images were
made of the Stokes $I\/$ emission.  We performed elliptical Gaussian
fits to the VLBA cores in the phase-referenced images, to determine
the galaxy positions listed in Table~1; those phase-referenced
positions carry the 2-D errors appearing in Table~2 for the appropriate
phase calibrator.  For adequately strong galaxies, we also performed
phase and amplitude self-calibrations on the (u,v) data and made
another set of images with higher dynamic ranges.  Figure~1 shows the
final images we obtained from the VLBA observations, at the
resolutions appearing in Table~1.  Three of these images (UGC\,01004,
UGC\,01413, and UGC\,08419) are phase-referenced, because the radio
flux densities for these galaxies are too weak ($<$ 10 mJy) for
self-calibration.  All other images in Figure~1 are self-calibrated.

For all source properties other than positions, the self-calibrated
VLBA image was used where available.  We performed Gaussian fits to
the radio core with the FWHM fixed to the angular resolution to
measure the peak core flux density, $S_p$.  The total flux density,
$S_t$, was measured by integration within the minimum region that
encloses the whole radio structure down to the lowest significant
contour level, \ie, three times the off-source noise level in the VLBA
image appearing in Table~1.  The results for $S_p$ and $S_t$ are
presented in Table~3 which also lists the peak core flux
density, $S_P$; the total radio flux density, $S_T$, and the
elongation position angle, $PA_J$, measured from VLA images presented
by Condon \& Broderick (1988).  Note that we use subscripts in lower
case for VLBA symbols and in upper case for VLA symbols.

Table~3 also lists the elongation position angles of the VLBA jets,
$PA_j$, determined as follows.  VLBA images were made with Gaussian
tapers close to the sizes of minor axes of the (u,v) coverage
ellipsoids and restored with circular Gaussian beams.  We first
determined the pixel position of the core in the image through a 2-D
Gaussian fit, and then made a circular cut (a ring) centered at the
core.  We then made a (1-D) Gaussian fit along the cut near the jet
direction to find the position angle where the intensity along the cut
is maximized.  The position angle of each jet was measured at
different distances from the VLBA core.  Since none of the VLBA jets
in our sample were found to be curved, the resultant position angles
were simply averaged and are listed in Table~3.  As a cross check, we
used the AIPS task IMFIT on images restored with the original
elliptical and with the circular Gaussian beams.  We found that the
elongation position angle of a VLBA jet agreed very well between the
two methods.
 
We also measured the brightness ratios between VLBA jets and counter
jets, also known as the sidedness ratio, $s$, by using a method
similar to that employed by Laing (1996).  We rotated the VLBA image
restored with a circular beam through 180\arcdeg~ about its core and
then divided that rotated image by the original image.  This yields a
sidedness image.  We have set the flux density cutoffs to be three
times the noise level in order to avoid meaningless values.  The
sidedness ratios presented in Table~3 are the mean value over each VLBA
jet, averaged within the region of the jet from 15 mas from the core to the
points where both side jets fade down to three times the noise level.  
We consider the ratios obtained through being divided by three
times the noise level as lower limits to the sidedness. 
If more than 50\% of the sidedness ratios are lower limits,
we consider the averaged value as a lower limit, as presented in Table~3.  
A similar analysis was performed on the VLA images, yielding the sidedness
ratios, $S$, appearing in Table~3.

\section{Results}\label{secresults}

We summarize here our main results from these VLBA observations.
First, as shown in Table~3, all sources are detected in these VLBA
observations.  UGC\,07654 is known to have VLBA jets as well.  Of the
17 sources, five (30\%) have unresolved radio cores, ten (60\%) show
core-jet structures, and two (10\%; UGC\,07360 and UGC\,11718) show
twin-jet structures.  A description of individual sources in the
Appendix~A provides more information.  Thus we show that parsec-scale
jets are common in low luminosity FR~Is with 12 sources with jets out of 17 
sources in a complete sample of 21.  Second, the flux densities in these 
VLBA sources are
dominated by their cores, $S_p$.  The VLBA jets, if detected, 
contribute a small (generally $<$ 20\%) fraction of the total radio
flux densities, $S_t$, in the VLBA images, and typically extend over
$\le$ 100 mas or 50 pc.  Third, as long as a VLBA jet is clearly
detected, it is nearly parallel to its VLA jet (Figure~2 and Table~3).
If the VLA jet is double-sided while the VLBA jet is only one-sided,
then the VLBA jet corresponds to the stronger side of the VLA jet.
Fourth, in all cases where the measurements of sidedness are without
ambiguity (\eg, where the VLBA sidedness ratio, $s$, is a lower limit
and larger than the measured VLA sidedness ratio, $S$; see Figure~3
and Table~3. Note that fewer measurements of sidedness ratio are presented
in Figure~3 since some sources have radio cores only so sidedness ratios
are not available), the sidedness ratios of the VLBA jets are always larger
than those of the corresponding VLA jets.  In the case that both
sidedness ratios are lower limits, the relationships are uncertain.
In UGC\,07360 where twin-jet structure is detected at both VLBA and
and VLA scales, the relation is slightly inverted, but this result is
not certain given the large errors.  Our results on VLBA-VLA jet
alignment and sidedness ratios are consistent with the results from a
study of FR~Is with more powerful radio cores (Lara \etal\ 1997).
Fifth, as long as a VLBA jet is clearly detected, it always shows some
internal structures, \ie, knots.  Comparing the radio images at
different scales, we notice that the structure of the knots displays
some self-similarity in their properties, \ie, the overall jet structures
are similar at different scales. For example, the spacing between knots
in the jets seems to scale with the resolution probed. 
Finally, all the VLBA jets
fade quickly down to noise levels.  The radio luminosity per unit
length of the jets, $L_j$, generally declines as $L_j \propto r^{-2}$,
where $r$ is the distance along the VLBA jets from their cores.
Using UGC\,00597 as an example, we illustrate in Figure~4 how the jet
position angle, surface brightness, and sidedness change with distance
from the core, both on VLBA scales (left panels) and on VLA scales
(right panels).  Detailed information for individual sources can be
found in Appendix~A.
Our results are generally consistent with those found in previous studies
of individual objects (e.g., Lara \etal\ 1997; Cotton \etal\ 1999). 

\section{Discussion}\label{secdis}

\subsection{Evidence For Doppler Boosting}\label{secdoppler}

The brightness distributions of the jets and counter jets are usually
very asymmetric, especially at parsec scales.  It has long been
suggested that this asymmetry is due to the Doppler beaming effect,
\ie, the two jets are intrinsically symmetric, but the brightness of
the jet moving towards the observer is boosted due to the relativistic
motion of the jet material, and that of the jet moving away from the
observer is dimmed.  The strongest evidence for this hypothesis is the
detection of superluminal motion in some radio galaxies and quasars
(\eg, Unwin \etal\ 1989). Recently, superluminal motion has been found
in some FR~I radio galaxies (1144+35 - Giovannini \etal\ 1999; NGC 315 - 
Cotton \etal\ 1999). This discovery directly supports the hypothesis 
of relativistic motion of the jet  (assuming cosmological distances).
Additional strong statistical evidence in support of the Doppler favoritism on 
large scales is the
discovery that the stronger jet is always on the side of the
less-depolarized lobe (Laing 1988; Garrington 1988).  This result
provides strong evidence that the brighter jet indeed points toward
the observer.  However, the Doppler beaming hypothesis is not the only
explanation that can account for the jet to counter-jet sidedness.
For example, an intrinsic difference in radio power of a pair of jets
can also be the reason why we see an asymmetric or one-sided jet with
VLBA observations.  One such model is the flip-flop jet scenario which
accounts for both the asymmetric parsec-scale jets and less asymmetric
kpc-scale jets (Rudnick \& Edgar 1984; Feretti \etal\ 1993).  Our
results support the Doppler favoritism hypothesis, because the
one-sided VLBA jet always corresponds to the stronger side of the VLA
jet and the sidedness of the VLBA jet is larger than that of the VLA
jet (see also, Parma \etal\ 1993).  These findings agree with the
prediction that the jet undergoes deceleration as it propagates
outwards (Bicknell 1994; Baum \etal\ 1997).  These facts are not
obviously predicted if we assume that the VLBA jet is intrinsically
one-sided.  For example, in the flip-flop model, whether the brighter
side of the parsec jet corresponds to the brighter side of the
kilo parsec jet depends on which side the parsec jet is ejected when
the observations are taken.

The Doppler beaming model is also supported by the following results.
We note that there are no extended structures associated with the VLBA
radio cores in some of our sample galaxies: UGC\,05073, UGC\,07455,
UGC\,08419, and (possibly) UGC\,12531.  At large scales, their jets
are more symmetric than those of the rest of our sample.  If the
symmetry of VLA-scale jets in these sources implies that their jets
lie close to the plane of the sky, rather than that the VLA-scale jets
are non-relativistic, then the lack of VLBA jets in these source is
consistent with the Doppler deboosting effect.  The deboosting factor is
$\delta ^{2+\alpha}$ where $\delta = \Gamma ^{-1} (1-\beta \cos \theta
)^{-1}$ (\eg, Zensus \& Pearson 1987), where $\delta$ is the jet
luminosity multiplier, $\alpha$ is the jet spectral index, $\beta=v/c$
is the jet speed, $\theta$ is the angle between the jet and the
line-of-sight, and $\Gamma$ is the bulk Lorentz factor of the jet.
For example, for $\Gamma \approx 5$ the jets are 60 times dimmer than
their rest frame brightness, assuming a spectral index $\alpha$=0.6
and $\theta=90\arcdeg$.  This factor of 60 would be sufficient to make the
majority of the observed VLBA jets in our sample undetectable if they were
to be placed in the plane of the sky.\footnote{The observed radio power
of the jets will be equal to the intrinsic power for an angle to the
line of sight of  $\approx$ 35\arcdeg~ for $\Gamma \approx 5$.}

We carried out calculations to simulate the above scenario, \ie, to
determine how the Doppler boosting and deboosting affect the statistical
properties of the observed radio structures.  We generated a set of
model two-sided radio sources with equal jet powers in their rest
frame.  The viewing angles of the two-sided sources were assumed to be
randomly distributed.  The logarithms of the power of radio jets at
parsec scale were assumed to populate a normal distribution.  The
Lorentz factors, $\Gamma$, were also assumed to satisfy a normal
distribution.  The distances to the sources were randomly assigned to
one of the distances from our sample.  We also set the brightness
detection limit to be 0.5 mJy~beam$^{-1}$, \ie, about three times the
noise level of our images.  Using the above assumptions, we found that
if we centered the normal distribution of the logarithm of radio jet
power at 29 (\ie, $10^{29}$ erg~s$^{-1}$~Hz$^{-1}$) with a standard
deviation 0.5, and centered the Lorentz factor at 5.0 with standard
deviation of 1.0, then we could roughly reproduce our observational
results.  (This value for the radio jet power corresponds to a source
with observed flux density 10 mJy at a distance of 100 Mpc, similar to
those in our sample [Table~3].)  Specifically, we found that about
17\% of the sources are expected to have two-sided jets, 55\% one-sided
jets, and 28\% no jet, similar to our observational results 
(cf: Sec.~\ref{secresults}).

Figure~5 illustrates how varying the parameters affects the statistical 
results. First, if we increase the width of the distribution of luminosity
we find fewer core-jet structures and more twin-jet structure and core
sources (panel {\em a\/}).  This occurs because the
Doppler effect tends to make one of the jets bright and the other one
dim (panel {\em d\/}).  If the instrumental detection limit lies (1)
between the flux densities of the pair of jets then we see a core-jet
structure, (2) below that of the weak jet then we see a twin-jet
structure, or (3) above the stronger jet then we then see no jet at
all.  Thus, a greater breadth to the distribution of  intrinsic jet luminosity
implies that more jets will be placed either above or below the detection 
limit, with the result that fewer sources are observed to have a core-jet 
structure.  
Second, our results are not very sensitive to the width of the distribution 
of Lorentz factor (panel {\em b\/}) because an increase in Lorentz factor 
tends to produce more core-jet and core sources while a decrease in 
Lorentz factor tends to produce more twin-jet sources, so these two effects 
somewhat balance out.  
Third, an increase in the median Lorentz factor will result in
more core jets and core only sources, and fewer twin-jet structures
(panel {\em c\/}).  This is understandable from panel {\em d\/} since
a larger Lorentz factor will increase the jet-to-counter jet ratio and
further deboost the jets near the plane of the sky, thereby producing
more core-jet structures and more undetectable jets but fewer twin-jet
sources.

\subsection{Why Do the VLBA Jets Fade?}\label{secfade}

While all the sources in our sample have kpc-scale radio jets, their
parsec-scale jets, if detected, all fade at a distance less than
$\sim$ 50 mas (or typically $\sim$ 30 pc) in our VLBA images.  For the
sources lacking VLBA jets, it is plausible that the jets fade at an
even shorter distance or are deboosted.  Is the fading of the jets in
the VLBA images an artifact related to the VLBA sensitivity?  The VLBA
sensitivity to large scale diffuse structure drops quickly at scales larger 
than 250 mas in our observations, which places all our observed jets 
well within the angular scales probed by our VLBA observations.  We 
assume the jets are continuous since we observe the jets at larger 
scales and the simple evolution of the sidedness ratio with distance suggests
the large scale jets are developed from the VLBA-scale jets we see here.  
The VLBA-scale jets in our images are laterally unresolved, as tested with 
Gaussian fits across the jets.  Thus the fading of a VLBA-scale jet is 
not due to the jet's being resolved out, thus lowering the peak surface 
brightness.  Rather, there is a real decrease of the observed jet power.

To quantify the fading of the jets, we performed a detailed analysis
of the evolution in jet brightness along the well defined jets in the
galaxies UGC\,00597, UGC\,01841, UGC\,06723, and UGC\,07360.  We
constructed diagrams of jet flux density per unit length \vs distance 
along the jet. 
We assume that the radio structure is made of three components
(an unresolved core, a jet, and a counter-jet) and we performed a
nonlinear fit to the data.  The basic strategy of the fit is as
follows.  We take a $\delta$-function as the core, plus two components
$\propto r^{-\lambda}$ as the jet and the counter jet.  We then
convolve these functions with a Gaussian and perform the nonlinear
fit.  The detailed equations for the fitting can be found in the
footnote under Table~4, which lists the fitted parameters.
We find values of $\lambda$ ranging from 1.6 to 2.3.  Jones \& Wehrle
(1997) presented multi-frequency VLBA observations for one of our
galaxies, UGC\,07360 and they found $\lambda \approx$ 2.0, slightly
steeper than our fitted value of 1.6.  However, their analysis
neglected the contamination of the jet from the core; if we fit the
jet without assuming the central $\delta$-function, we do reproduce
their $\lambda \approx 2.0$.

It is worth noting that, in the above discussion, the fading of the
jets only applies to scales of parsecs to tens of parsecs.  At scales
beyond hundreds of parsecs the jets appear to evolve differently,
probably due to the interaction between the jets and the external
medium.  In fact, the evolutionary slope $\lambda$ in UGC\,06723 is
about 1.0 for the jet between 100 and 400 pc (Figure~13{\em a} in Baum
\etal\ 1997), although that slope is based on the variation of the
peak jet amplitude with distance along the jet.  Another interesting
source is UGC\,03087 (3C\,120), for which Walker \etal\ (1987) discuss
radio data on the jet ranging from 0.5 pc to over 400 kpc (for $H_0$ =
100) and found an overall slope of 1.27.  However, we note that in
their Figure~15 the slope within 100 pc is about 2.3, steeper than its
overall value and more in agreement with our results on these size
scales.

The fading of the jets on parsec scales could be the result of at
least three mechanisms or their combination: (1) The jet is
decelerated from ultra-relativistic to intermediate relativistic
speed, thereby effectively reducing the Doppler beaming effect.  (2)
The jet power is reduced through synchrotron losses.  (3) The jet
undergoes adiabatic expansion which reduces the jet internal energy.
Each of these mechanisms will now be considered in turn.

\subsubsection{Doppler Boosting}\label{secsubdoppler}

Here we consider whether the decrease in jet brightness could be
caused by a reduction in Doppler boosting as an initially relativistic
jet decelerates. An important observational constraint is that
the two-sided and one-sided jets fade in the same manner. 

According to Baum \etal\ (1997,1998), the observed jet surface brightness
variation satisfies 
\begin{equation}
I_j \propto (\Gamma \beta )^{-(2\alpha + 3)/3}
\delta ^{2+\alpha} w^{-(10\alpha + 9)/3} \propto \Gamma^{-4.1}
\beta^{-1.4} w^{-5.2} (1-\beta \cos \theta)^{-2.7}
\end{equation}
for jets with magnetic fields running parallel to the jet axis, and
\begin{equation}
I_j \propto (\Gamma \beta)^{-(5\alpha + 6)/3} \delta ^{2+\alpha}
w^{-(7\alpha + 6)/3} \propto \Gamma^{-5.7} \beta^{-3.1} w^{-3.5}
(1-\beta \cos \theta)^{-2.7}
\end{equation}
for jets with magnetic fields running perpendicular to the jet axis.  
Hence the variation of the laterally integrated brightness 
goes as 
\begin{equation}
I_j \propto \Gamma^{-4.1} \beta^{-1.4} w^{-4.2} 
(1-\beta \cos \theta)^{-2.7}
\end{equation}
 for the $B_\parallel$ case and 
\begin{equation}
I_j \propto \Gamma^{-5.7} \beta^{-3.1} w^{-2.5} (1-\beta \cos \theta)^{-2.7}
\end{equation}
for the $B_\perp$  case.  Here $\Gamma$
is the jet Lorentz factor, $w$ is the transverse jet width, $\theta$
is angle to the line-of-sight, $\alpha=0.65$ is the jet spectral index, and
$\delta = \Gamma ^{-1} (1-\beta \cos \theta )^{-1}$.  
 We calculate the jet brightness evolution
for different cases, \ie, different lines of sight, different jet
opening angles  (with adiabatic effects taken into account)
 and different magnetic field configurations and the results
are shown on Figure~7. The jet brightness $I_j$ is plotted against
jet velocity $\beta$ where we assume that the jet is decelerated 
linearly with the distance from the core over these scales. 
(Note: since the jets are laterally unresolved, brightness $I_j$ and 
luminosity per unit length $L_j$ are linearly associated; so this plot 
can be considered equivalent to a plot of  $L_j$ against $\beta$ 
and thus can be directly compared with our observational results.)

Figure~7 shows that the behavior of the jet brightness as the jets decelerate
is somewhat complicated because there are two competing effects. 
(1) The change in the Doppler boosting factor as the jets decelerate causes 
the jets pointing towards us to dim and the jets pointing away from us
and those near the plane of the sky to brighten. 
(2) The deceleration of the jet compresses the jet fluid along the jet axis
and increases the jet density and the magnetic field. This causes the 
jets to brighten. The jets dominated by perpendicular magnetic field brighten 
much more than the jets dominated by parallel magnetic field since 
 $B_\parallel \propto w^{-2}$ while  $B_\perp \propto (w \beta \Gamma) ^{-1}$
(Baum \etal\ 1997,1998). 

The fact that model jets pointed towards us evolve differently than those
 in the plane of the sky is inconsistent with the result that the
one-sided and two-sided jets are observed to fade in  the same manner.
The fact that over much of the orientation range, the model jets brighten
rather than dim is inconsistent with the fact that the jets are
observed to fade. Thus, we can rule out the hypothesis that deceleration 
of an initially relativistic jet is responsible for the observed fading
of the jets on the tens of pc scale.

\subsubsection{Synchrotron Losses}\label{secsyn}

Here we consider the hypothesis that synchrotron losses can  
account for the jet fading. 
 These  sources have equipartition magnetic fields of $\sim
5\times 10^{-3}$ G (Table~4).  Using the standard formula $t_s =
8\times 10^8 B^{-2} \gamma^{-1}$ s to calculate the synchrotron
lifetime and assuming an electron (not bulk) Lorentz factor $\gamma
\sim 10^3$, which is typical for GHz radio emission in the above
magnetic field, we find $t_s \sim$ 1000 yr.  Thus, the corresponding
travel distance of an electron in the jet is about 300 pc for a jet
with speed c.  This value is more than 10 times the scale of fading
seen in our VLBA images.  However, the magnetic field could be
stronger than the equipartition field on these scales.  Increasing the
magnetic field by a factor of 10 above equipartition reduces the
electron lifetime by a factor of order 100 and the fading would thus
occur on the observed scales.  In fact, Venturi \etal\ (1993)
estimated a higher value of the brightness temperature for the jets in
UGC\,00597 than we do, implying a higher equipartition magnetic field
and thus a shorter life for synchrotron-emitting electrons.  Jones \&
Wehrle (1997) found that the spectral indices of the jets in
UGC\,07360 steepen with the distance from the core, as expected for
the synchrotron loss mechanism.  Thus, it is possible that the fading
of a VLBA jet is due simply to synchrotron losses in a magnetic field
about one order of magnitude higher than the equipartition value.

We assume (1)  the jet undergoes virtually no expansion, 
consistent with observations of M 87 on these scales (Junor \etal\
1999; Junor \& Biretta 1995),
(2) synchrotron 
radiation is the dominant energy loss mechanism for the ultra-relativistic
electrons, (3) the jet has constant velocity $\approx c$,  and (4)
constant magnetic field.
The latter is justified because the sound speed of the jet is much
smaller than that of light speed as implied from the brightness
temperature in Table~4.  The energy change of a synchrotron emitting
electron satisfies $dE/dt = -\eta E^2$, where $\eta=1.6\times 10^{-3}
B^2$ and B is the magnetic field.  If we assume the distribution of
synchrotron electrons is $N(E,0)=N E^{-p}$, where $p = (2\alpha + 1)$,
we then have $N(E,t) = N E^{-p} (1- \eta E t)^{(p-2)}$ at any later
time (Kardashev 1962; Wilson 1975).  Note here we have $E < 1/\eta t$,
\ie, there are no electrons with energy above $1/\eta t$ because of 
synchrotron losses.  Thus, we have  
\begin{equation}
L_j \propto (1- \eta E t)^{(p-2)}
\propto (1- \eta E r/c)^{(p-2)} \propto (1-A r)^{(2\alpha-1)}
\end{equation}
as a result of constant jet speed, where $L_j$ is the radio
luminosity per unit length of the jets as mentioned in Sec.~\ref{secresults}.
This relationship is not a simple
power-law and is effectively flatter ($\propto -r$ near the origin)
than observed (Figure~6), \ie, it is not an acceptable fit to the
observed jets (see Table~4).  Thus, we rule out the possibility that a
simple synchrotron loss model can explain the fading of the VLBA jets
because (1) it would require a magnetic field an order of magnitude
higher than the equipartition value and (2) the observed decline of
jet brightness is inconsistent with the predicted slope. 

\subsubsection{Adiabatic Expansion of a Constant Velocity Jet}\label{secexpan}

If adiabatic evolution is the dominant energy loss mechanism in the
jet, we expect that the jet luminosity per unit length will depend on
its transverse width $w$ as $L_j \propto w^{-\lambda}$ and will depend
on distance along the jet $r$ as $L_j \propto (w_0 + 2 r \Theta)^{-\lambda}$.  
These dependencies assume a constant jet opening
angle, $\Theta$, and constant jet velocity, $\beta$.  Moreover, for an
optically thin jet with spectral index 0.65, $\lambda$ will be 4.2,
2.5 or 3.1, depending on whether the magnetic field runs parallel to
the jet, runs perpendicular to the jet, or is maintained at the
equipartition value (Jones \& Wehrle 1997).

Very little data are presently
available on the magnetic field structures in FR~I radio galaxies on
these linear scales.  We note that powerful quasars tend to have magnetic 
fields parallel to the jets on
scales of parsecs to tens of parsecs, while those in BL Lacertae
objects tend to be perpendicular to the jets (Cawthorne \etal\ 1993;
Gabuzda \etal\ 1994; Wardle 1998).  If we assume that the current BL
Lac/FR~I unification schemes are correct, then the magnetic fields in
FR~Is should also be perpendicular to the jets.  
M~87 is a counter example to this extrapolation and shows magnetic
field mostly parallel to the jets except at the knots (Perlman \etal\
1999). This question can only be answered with polarimetry of the
jets in our sample.   A perpendicular
magnetic field would produce an exponent of 2.5 for the variation of
jet luminosity per unit length \vs distance, which is comparable to
the observed value of 2.0 (Table~4).  

In summary, we find that adiabatic expansion of a constant velocity jet
is consistent with the observed jet fading under the assumption that 
the magnetic field is perpendicular to the jet axis. 
This prediction should be tested with
(1) VLBA polarimetry to determine the magnetic field 
morphologies and (2) higher resolution observations to determine whether 
the jets are  indeed expanding. An important implication of this result
is that the deceleration of these jets which is implied by the Unified Scheme
does not take place on tens of pc scales, but must instead occur
on larger scales (\eg, hundreds of parsecs or larger). 
Baum \etal\ (1997) showed that the properties of the jet in 3C 264 are 
consistent with strong deceleration on hundreds of pc scales. Laing
\etal\ (1999) argue for deceleration on similar or slightly larger
scales in a sample of B2 radio galaxies.  
 
\section{Summary}\label{secsumm}

We present VLBA observations of 17 members of a complete sample of 21 nearby
FR~I radio galaxies for which we also have VLA, HST, and ROSAT data.
We detected parsec-scale radio emission in all 17 sources observed
with the VLBA.  Five VLBA sources show only a radio cores, ten sources
show core-jet structures, and two sources show twin-jet structures.
We find that the VLBA images are core-dominated, while the VLBA scale
jets, if detected, contribute only a small fraction (mostly $<$ 20\%)
of the total VLBA radio flux densities.  We also find that (1) all
VLBA jets are aligned with VLA jets; (2) the jet-to-counter-jet
sidedness ratio measured with the VLBA is generally larger than that
measured with the VLA; (3) the VLBA jets fade with distance from the
AGN core as luminosity per unit length $L_j \propto r^{-2.0}$; (4) the
observations suggest self-similarity in the structures of the knots.
We argue that results (1) and (2) are consistent with Doppler boosting
effects on the parsec scales and deceleration of the jets between
scales of tens of parsecs and kilo parsecs.  We show that a
distribution of bulk Lorentz factors centered near $\Gamma = 5$ can
reproduce our VLBA detection statistics for core, core-jet, and
twin-jet sources.  We consider three hypotheses to explain our result 
that the jets fade on the  tens-of-parsecs scale.  (i) If the
fading is due to a decrease of the Doppler boosting as the jet
decelerates, it would predict very different behavior for the
one-sided and two-sided jets, contrary to observations.  Also, in 
most cases we would expect the jets to {\it brighten}, not fade, due
to compression of the jet fluid as the jet decelerates. (ii)
Synchrotron losses in a magnetic field about an order of magnitude
higher than our estimated equipartition values would produce dimming
on the appropriate size scales, but with a slope to the brightness
evolution which is inconsistent with the observed value.  
(ii) Expansion losses in an adiabatically expanding jet, with constant 
velocity and opening angle, are roughly consistent with the observations, 
provided the magnetic field structure in the jets is perpendicular to 
the jet axis.  This should be tested with VLBA polarimetry to determine
the magnetic field structure and higher resolution observations to determine
the jet expansion. Our results imply  that the jets do not
decelerate on the tens of pc scale but must decelerate on larger scales. 

\acknowledgements C.X.\ is grateful for the hospitality of NRAO, would
like to thank Alan Roy for help with some of the VLBA data reduction,
and was partially supported by a grant from the STScI Director's
Discretionary Research Fund. We are grateful to the anonymous referee 
for helpful comments on the paper.

\clearpage

\appendix
\section{Comments on Individual Sources}\label{secappen}

\noindent {\bf UGC\,00408 (NGC\,193):} The VLBA image of this source
shows clear core-jet radio structure.  There is also evidence for a
knot with flux density 1.4 mJy, located 42.3 mas from the core at
position angle 104\arcdeg.  This knot is thus on the jet side, yet well
beyond the faded jet (Figure~1).  X-ray observations using the High
Resolution Imager on board the ROSAT observatory reveal that the X-ray
structure of this source consists of two components: an east-west
component elongated near the jet direction and lying on the optical
center, plus an elliptical ring oriented north-south and matching the
size of the optical galaxy (Xu \etal\ 2000).

\noindent {\bf UGC\,00597 (NGC\,315):} This source has been
extensively studied using the VLA and VLBI (\eg, Venturi \etal\ [1993],
Cotton \etal\ [1999] and references therein). The radio jets are 
found to have large sidedness ratios from mas to arcmin scales.

\noindent {\bf UGC\,00689 (NGC\,383 = 3C\,31):} On the arcsec scale,
this source displays an S-type symmetry and a sidedness ratio of $\ge$
20.  The sidedness ratio of our VLBA image is $\ge$ 6.  Lara \etal\
(1997) measured sidedness ratios at 30\arcsec, 15\arcsec, and 8 mas to
be 1.5, 6, and $>$16, respectively, with VLA and VLBI data at 5 GHz.
The spectral index of the VLBI core between 1.65 and 4.95 GHz is found
to be -0.44, indicating an inverted spectrum and assuming no flux density
variations between our observations on 1997 April 9 and those of Lara
\etal\ (1997) on 1993 February 25.

\noindent {\bf UGC\,01004 (NGC\,541):} A weak radio core was detected
in this source, which may be marginally resolved in our VLBA image.

\noindent {\bf UGC\,01413 (NGC\,741 = 4C\,05.10):} This galaxy is a
member of a pair of binary galaxies NGC 741/742.  The VLA image
displays a core residing near the center of NGC\,741 plus a knot
30\arcsec~ to the east of that core (Birkinshaw \& Davis 1985). Our
VLBA image marginally detects a core with no obvious extended structure
associated with it.

\noindent {\bf UGC\,01841 (3C\,66B):} This source displays a C-type
symmetry in its radio jets on arcmin scales (\eg, Hardcastle \etal\
1996).  Optical synchrotron emission associated with the radio jets
has also been detected in this source (Butcher \etal\ 1980).  Our VLBA
image shows a core-jet structure (Figure~1).  The lower resolution
image restored with a circular Gaussian beam (cf: Sec.~\ref{secdata})
reveals a knot at 59 mas to the north-east of the core.  The latter
image resembles the EVN-MERLIN image made by Fraix-Burnet \etal\
(1997) except that the overall scale of our image is larger,
suggesting self-similarity between small and intermediate scale
structure.

\noindent {\bf UGC\,03695 (NGC\,2329):} This galaxy is a cluster
member in Abell 569.  It hosts a wide angle tailed radio source with
the bulk of the radio emission coming from the core (Feretti \etal\
1985).  The VLBA detects a core-jet elongated in the direction
corresponding to the northeastern radio tail.

\noindent {\bf UGC\,05073 (NGC\,2892):} An unresolved radio core was
detected in our VLBA image, with no associated extended structure.
The lower limit of the core to jet flux ratio is estimated to be $\sim$ 140.
 
\noindent {\bf UGC\,06723 (NGC\,3862 = 3C\,264):} The overall radio
structure of this galaxy is very unusual.  It consists of three
structural components: a small-diameter core, a one-sided jet, and a
large amorphous emission plateau (Bridle \&Vall\'ee 1981).  Optical
synchrotron emission is associated with the jet (Crane \etal\ 1993;
Baum \etal\ 1997).  Recent global VLBI and VLA images of this source
can be found in Lara \etal\ (1997), while MERLIN images appear in Baum
\etal\ (1997).

\noindent {\bf UGC\,07360 (NGC\,4261 = 3C\,270):} HST observations of
this galaxy reveal a nuclear dust disk which is nearly perpendicular
to the large scale radio jets (Jaffe \etal\ 1993, 1996).  An enclosed
mass of $\sim 4\times 10^7~M_{\sun}$ within 0.1 pc was also suggested
by those studies.  Our VLBA image shows this source has a twin-jet
structure at parsec scales.  Parsec-scale twin jets have been detected
in fewer than ten FR~I radio sources so far (Xu \etal\ 1999).
Parsec-scale images obtained with the VLBA at at other wavelengths can
be found in Jones \& Wehrle (1997).

\noindent {\bf UGC\,07455 (NGC\,4335):} This source has well-defined,
two-sided jets on scales of tens of arc seconds. However, we are unable to
detect any jet-like structures associated with the nearly unresolved
radio core at mas scales.  The VLBA core to  jet flux density ratio is
estimated to be greater than $\sim$ 23.

\noindent {\bf UGC\,07494 (NGC\,4374 = 3C\,272.1 = M84):} We detect
core-jet structure in this source, with the VLBA jet pointing towards
the northern and the stronger and narrower of the VLA jets.

\noindent {\bf UGC\,08419 (NGC\,5127):} This source has quite
symmetric two-sided jets on scales of tens of arc seconds.  Yet the
VLBA image shows no evidence for jets.   The core to 
jet flux density ratio is estimated to be greater than about 10.

\noindent {\bf UGC\,08433 (NGC\,5141):} Core-jet structure is detected,
with the mas jet pointing to the southern and slightly stronger VLA
jet.

\noindent {\bf UGC\,09058 (NGC\,5490):} Similar to the situation in
UGC\,08433, a core-jet structure is detected with the mas jet pointing
to the eastern  and slightly stronger VLA jet.

\noindent {\bf UGC\,11718 (NGC\,7052):} We see evidence for two-sided
emission (in NNE and SSW directions) in this source at mas scales, 
with the stronger side pointing
to the nearly one-sided jet on scales of tens of arc seconds.  If
confirmed, this will be a very special case of twin jets detected at
VLBA scales but one jet detected at VLA scales.  But we should be very
cautious about the possible twin-jet structure since it could also be
a symmetrization artifact due to inaccurate amplitude calibration.  We
are unable to perform any further tests on this feature due to its
weakness.  However, we still count this source as a twin-jet source
because the detected twin-jet feature is well aligned with the large
scale jet.
  
\noindent {\bf UGC\,12531 (NGC\,7626):} We detect weak evidence for a
core-jet structure in this source.  The jet is misaligned by
10\arcdeg~ from the direction of the northern VLA jet.  This apparent
misalignment could be a real offset between VLBA and VLA jets.  However
due to the faintness of the jet, the measurement is uncertain.

\newpage 

\centerline{\bf Figure Captions}
\oddsidemargin  0.2in
\parindent -0.4in

{\bf Fig~1.} VLBA images of Stokes $I\/$ emission at 1.67 GHz from 17
FR~I radio galaxies.  The lowest contour levels correspond to three
times the noise levels and contours are spaced at 
3$\sigma\times$ [-3,-2,-1,1,2,4,8,16,32,64,128,256,512]. The images of
UGC\,00408, UGC\,01413, and UGC\,08419 are phase-referenced.  All
other images are self-calibrated.

{\bf Fig~2.} Histogram of the position angle offsets between jets on
VLBA (mas/pc) and VLA (arcsec/kpc) scales. This plot indicates
that the jets at VLBA scale and those at VLA scale  are very well
aligned.

{\bf Fig~3.} Jet sidedness at both VLBA (mas/pc) and VLA (arcsec/kpc) 
scales. The arrow-like bars associated with the data represent
the lower/upper limits. The dashed line indicates equal jet 
sidedness on the VLBA and VLA scales. The fact that more points
lie above the dashed line indicates that the sidedness at VLBA
scales are on average larger than those at VLA scales, which further
implies a deceleration of the radio jets. 

{\bf Fig~4.} UGC~00597. Variation of jet position angle, surface 
brightness, and sidedness ratio as a function of distance from the 
core on both VLBA (mas/pc) and VLA (arcsec/kpc) scales.

{\bf Fig~5.}  Panels (a), (b), and (c) indicate the fractions of
sources with two detectable jets (above the open circles), only one
detectable jet (between the open and filled circles), or core only
(below the filled circles) as functions of (a) the standard deviation of
the distribution of the jet luminosity, (b) standard deviation of
the distribution of the jet bulk Lorentz factor, and (c) the
mean jet bulk Lorentz factor.  Panel (d) shows the Doppler boosting factors
of jets (solid lines and upper curves) and counter jets (dotted lines and
lower curves) as a function of viewing angle for bulk Lorentz factors 
2, 3, 4, 5, and 6. The open and filled circles in panel (a), (b) and (c)
were obtained through simulations and the curves running through the dots
are simply fits to the simulated results. The curve for Lorentz factor equals
5 in panel (d) is plotted with a thicker line for clarity.

{\bf Fig~6.} Fits to the radio jet brightness as a function of distance 
from the core for four sources where the VLBA jets are clearly detected.
The solid points in the upper part of each plot represent observed
data. The solid line is the fit to the data with  a power law 
and the dash-dotted line is the fit with a synchrotron loss model. 
The solid points in the lower part of the plot represent the residuals 
from the power law fit and the open circles represent those 
from the  synchrotron loss model (cf: Table 4).

{\bf Fig~7.} Illustration of the predicted variation of jet brightness $I_j$ 
with jet bulk velocity in a decelerating relativistic jet 
due to both Doppler boosting and adiabatic expansion/compression 
effects (see Sec.~\ref{secsubdoppler}). The expected variations are shown 
for a range of line-of-sight inclination angles, three values of 
jet opening angle (5\arcdeg, 15\arcdeg, 25\arcdeg),
and magnetic field oriented both parallel and perpendicular to the jet axis. 
In each plot, the curves from top to bottom corresponds to jets at
line-of-sight inclination angles 5\arcdeg, 15\arcdeg, 30\arcdeg,
45\arcdeg, 60\arcdeg, 75\arcdeg, 90\arcdeg, 120\arcdeg and
150\arcdeg, with the 90\arcdeg curves plotted with a thicker line for clarity.

\newpage

\begin{center}
{\sc Table 1. VLBA Positions, Resolutions, and Sensitivities for
              UGC Galaxies}
\end{center}
{\small
\begin{center}
\begin{tabular}{ccclcllcc} \hline \hline
\multicolumn{1}{c}{}&
\multicolumn{1}{c}{Other}&
\multicolumn{1}{c}{$m_p$}&
\multicolumn{1}{c}{Hubble}&
\multicolumn{1}{c}{$D_{75}$}&
\multicolumn{1}{c}{}&
\multicolumn{1}{c}{}&
\multicolumn{1}{c}{Resolution}&
\multicolumn{1}{c}{Image $r.m.s.$} \\
UGC& name & (mag)& Type& (Mpc)& ~~~~~$\alpha_{J2000}$& ~~~~~$\delta_{J2000}$& 
(mas,mas,\arcdeg)& (mJy/b.a.) \\ \hline
00408&NGC 193  & 13.9& S0  & 60.3& 00 39 18.5829&    +03 19 52.584&     9.8,3.8,$-$6 & 0.10 \\
00597&NGC 315  & 12.5& E   & 69.1& 00 57 48.8834$^a$&+30 21 08.812$^a$&10.3,3.9,$-$7 & 0.11 \\
00689&3C 31    & 13.6& E0  & 71.1& 01 07 24.9593&    +32 24 45.237&    10.2,3.9,$-$7 & 0.12 \\
01004&NGC 541  & 14.0& E   & 73.8& 01 25 44.3078&  $-$01 22 46.522&    12.4.5.6,   0 & 0.10 \\
01413&NGC 741  & 13.2& E   & 76.3& 01 56 20.9902&    +05 37 44.277&    11.0,4.3,$-$5 & 0.16 \\
01841&3C 66B   & 15.0& E   & 88.1& 02 23 11.4073&    +42 59 31.403&     9.4,3.9,$-$8 & 0.13 \\
03695&NGC 2329 & 13.7& E-S0& 78.3& 07 09 08.0061&    +48 36 55.733&    11.2,3.6,$-$10& 0.13 \\
05073&NGC 2892 & 14.4& E   & 90.8& 09 32 52.9316&    +67 37 02.630&    10.9,3.8,   21& 0.12 \\
06635&NGC 3801 & 13.3& E   & 42.0& 11 40 17.31  &    +17 43 36.8  &         ---      &  --- \\
06723&3C 264   & 14.0& E   & 84.9& 11 45 05.0099&    +19 36 22.756&    10.3,4.1,$-$4 & 0.11 \\
07115&         & 14.4& E   & 89.5& 12 08 05.81  &    +25 14 16.4  &         ---      &  --- \\
07360&3C 270   & 12.0& E   & 27.7& 12 19 23.2162&    +05 49 29.702&    10.6,4.5,$-$2 & 0.10 \\
07455&NGC 4335 & 13.7& E   & 63.1& 12 23 01.8881&    +58 26 40.384&     8.0,4.4,$-$4 & 0.14 \\
07494&M 84     & 10.8& S0  & 14.6& 12 25 03.7433$^a$&+12 53 13.143$^a$&10.8,4.0,$-$5 & 0.15 \\
07654&M 87     & 10.4& E   & 16.0& 12 30 49.4234$^b$&+12 23 28.044$^b$&     ---      &  --- \\
08419&NGC 5127 & 13.9& E   & 64.3& 13 23 45.0156&    +31 33 56.703&    10.1,4.2,$-$17& 0.12 \\
08433&NGC 5141 & 13.9& S0  & 70.0& 13 24 51.4403&    +36 22 42.763&     9.8,4.1,$-$19& 0.12 \\
09058&NGC 5490 & 13.4& E   & 66.2& 14 09 57.2984&    +17 32 43.911&    10.4,4.2,$-$8 & 0.12 \\
11718&NGC 7052 & 14.0& E   & 69.1& 21 18 33.0446&    +26 26 49.251&    10.4,3.8,$-$6 & 0.10 \\
12064&3C 449   & 14.6& E-S0& 72.4& 22 31 21.35  &    +39 21 33.2  &         ---&  --- \\
12531&NGC 7626 & 12.8& E   & 48.3& 23 20 42.5391&    +08 13 00.992&    11.9,5.1,   6 & 0.18 \\
\hline \hline
\end{tabular}
\end{center}
}
Notes: $^a$ M.\ Eubanks, private communication.
       $^b$ Ma \etal\ 1998.

\newpage

\begin{center}
{\sc Table 2. VLBA Phase Calibrators for UGC Galaxies}
\end{center}
{\small
\begin{center}
\begin{tabular}{cccccl} \hline \hline
\multicolumn{1}{c}{}&
\multicolumn{1}{c}{Phase}&
\multicolumn{1}{c}{}&
\multicolumn{1}{c}{}&
\multicolumn{1}{c}{2-D Error}&
\multicolumn{1}{c}{}\\
UGC&Calibrator&$\alpha_{J2000}$&$\delta_{J2000}$&(mas)& Ref.\\ \hline
00408& J0049+0237& 00 49 43.2363& +02 37 03.785& 14 &   1,3 \\
00597& J0112+3208& 01 12 50.3276& +32 08 17.548& $>$55& 1,3 \\
00689& J0112+3208& 01 12 50.3276& +32 08 17.548& $>$55& 1,4 \\
01004& J0127+0158& 01 27 22.8822& +01 58 24.534& 14 &   3   \\
01413& J0201+0343& 02 01 51.5093& +03 43 09.256& 14 &   1,3 \\
01841& J0230+4032& 02 30 45.7068& +40 32 53.087& 12 &   1,2 \\
03695& J0710+4732& 07 10 46.1052& +47 32 11.142& 12 &   1,2 \\
05073& J0903+6757& 09 03 53.1559& +67 57 22.683& 12 &   1,2 \\
06723& J1148+1840& 11 48 37.7776& +18 40 08.983& 14 &   1,3 \\
07360& J1222+0413& 12 22 22.5501& +04 13 15.780& 14 &   1,3 \\
07455& J1217+5835& 12 17 11.0203& +58 35 26.228& 12 &   1,2 \\
07494& J1213+1307& 12 13 32.1412& +13 07 20.373& ---&   1   \\
08419& J1329+3154& 13 29 52.8650& +31 54 11.047& 55 &   1,4 \\
08433& J1317+3425& 13 17 36.4935& +34 25 15.923& 55 &   1,4 \\
09058& J1412+1334& 14 12 36.3728& +13 34 38.166& 14 &   1,3 \\
11718& J2114+2832& 21 14 58.3340& +28 32 57.206& 55 &   1,4 \\
\hline \hline
\end{tabular}
\end{center}
}
References: 1.\ Peck \& Beasley 1998. 
            2.\ Patnaik \etal\ 1992.
            3.\ Browne \etal\ 1998.
            4.\ Wilkinson \etal\ 1998.

\newpage

\oddsidemargin -1cm
\begin{center}
{\sc Table 3. Source Properties in the VLBA (1.67 GHz) and VLA (1.49 GHz) Images}
\end{center}
{\small
\begin{center}
\begin{tabular}{ccccccccccc} \hline \hline
\multicolumn{1}{c}{}&
\multicolumn{2}{c}{VLBA Flux Density}&
\multicolumn{2}{c}{VLA Flux Density}&
\multicolumn{2}{c}{Morphology$^a$}&
\multicolumn{2}{c}{Position Angle}&
\multicolumn{2}{c}{Sidedness Ratio} \\
   &     Peak &Total&     Peak &Total& VLBA&VLA& VLBA&VLA&   VLBA&VLA \\ 
   &     $S_p$&$S_t$&     $S_P$&$S_T$& & &  $PA_j$&$PA_J$&    $s$&$S$ \\
UGC&(mJy/b.a.)&(mJy)&(mJy/b.a.)&(mJy)& & &(\arcdeg)&(\arcdeg)&   &    \\ 
\hline
00408& 29.8$\pm$0.5  & 36.6$\pm$0.6  & 40$\pm$0.9  & 145$\pm$1.7  & CJ& CJ&   101&    92&   $>$3&     $>$7 \\
00597&223.7$\pm$19.0 &356.0$\pm$24.0 &396$\pm$2.1  & 501$\pm$2.4  & CJ& CJ& $-$50& $-$49&  $>$25&     $>$4 \\
00689& 44.1$\pm$1.5  & 51.6$\pm$1.6  & 89$\pm$1.4  & 198$\pm$2.1  & CJ& TJ& $-$18& $-$21&   $>$6&    $>$20 \\
01004&  1.9$\pm$0.4  &  2.7$\pm$0.4  &  8$\pm$0.8  &  33$\pm$1.6  &  C& CJ&   ---&$-$104&    ---&      --- \\
01413&  4.4$\pm$0.6  &  8.6$\pm$0.8  & 13$\pm$1.0  & 104$\pm$2.8  &  C& CJ&   ---&   ---&    ---&      --- \\
01841&112.0$\pm$5.0  &157.4$\pm$6.0  &131$\pm$12.0 & 510$\pm$24.0 & CJ& CJ&    50&    50&  $>$25&    $>$20 \\
03695& 49.7$\pm$1.6  & 59.1$\pm$1.8  &117$\pm$2.2  & 271$\pm$3.3  & CJ& CJ& $-$30& $-$29&   $>$3&      --- \\
05073& 15.3$\pm$0.6  & 17.5$\pm$0.6  & 22$\pm$0.7  &  94$\pm$1.4  &  C& TJ&   ---&    52&    ---&$\sim$1.5 \\
06635&            ---&            ---&          ---&1032$\pm$21.0 &---& TJ&   ---&   115&    ---&$\sim$1.8 \\
06723&122.9$\pm$6.1  &166.1$\pm$7.1  &386$\pm$9.3  & 848$\pm$14.0 & CJ& CJ&    30&    25&  $>$12&     $>$8 \\
07115&            ---&            ---&          ---&  41$\pm$2.0  &---& CJ&   ---&   116&    ---&   $>$2.5 \\
07360& 67.4$\pm$9.3  &160.4$\pm$14.0 &165$\pm$1.0  & 173$\pm$1.1  & TJ& TJ& $-$93& $-$96&$\sim$1&$\sim$1.5 \\
07455&  9.2$\pm$0.5  & 10.9$\pm$0.5  & 15$\pm$1.5  & 115$\pm$4.2  &  C& TJ&   ---&    79&    ---&$\sim$1.0 \\
07494&106.0$\pm$1.6  &119.0$\pm$1.7  &112$\pm$13.0 &2254$\pm$58.0 & CJ& TJ&     1&     0&   $>$5&$\sim$1.5 \\
07654&         1570.0&            ---& 3600$\pm$200&56090$\pm$150 & CJ& CJ&$-84^b$&$-70^b$ &  $>200^c$& $>$15  \\
08419&  3.9$\pm$0.5  &  5.2$\pm$0.6  &  7$\pm$0.7  &  41$\pm$1.7  &  C& TJ&   ---&   118&    ---&$\sim$1.0 \\
08433& 35.3$\pm$2.0  & 43.8$\pm$2.2  & 71$\pm$2.7  & 859$\pm$9.4  & CJ& TJ&$-$168&$-$170&   $>$3&$\sim$1.5 \\
09058& 20.0$\pm$1.0  & 25.5$\pm$1.1  & 41$\pm$7.0  & 403$\pm$22.0 & CJ& TJ&    75&    75&   $>$2&$\sim$1.0 \\
11718& 27.9$\pm$0.9  & 32.1$\pm$1.0  & 36$\pm$0.9  &  98$\pm$1.5  & TJ& CJ&$-$157&$-$158&   $>$2&      --- \\
12064&            ---&            ---&          ---&  18$\pm$1.1  &---& TJ&   ---&    11&    ---&      --- \\
12531& 12.5$\pm$0.6  & 15.4$\pm$0.7  & 23$\pm$1.0  & 128$\pm$0.4  & CJ& TJ&    44&    32& $>$1.5&$\sim$1.0 \\
\hline \hline
\end{tabular}
\end{center}
}
Notes: $^a$ C means core, CJ means core-jet, TJ means twin-jet. 
       $^b$ Values from Junor \& Biretta 1995.
       $^c$ Values from Reid \etal\  1989.
       VLBA and VLA values are denoted by lower case and upper case
       subscripts, respectively. 
\newpage

\oddsidemargin 0in
\begin{center}
{\sc Table 4. Fits to VLBA Jet Evolution}
\end{center}
{\small
\begin{center}
\begin{tabular}{crrrccccc} \hline \hline
UGC& $A_1$& $A_2$& $A_3$& $\lambda$& $T_b$ (K)& $B_{eq}$ (G)&
${\chi_p}^2$& ${\chi_s}^2$ \\ \hline
00597& 34.3& 97.4& 18.6& 2.1& $8.2\times10^7$& $6.6\times10^{-3}$& 9.9& 150 \\
01841& 16.6& 41.9&  3.6& 2.2& $4.5\times10^7$& $5.5\times10^{-3}$& 4.4& 100 \\
06723& 17.1& 67.6&  9.4& 2.3& $3.8\times10^7$& $5.3\times10^{-3}$& 3.9&  33 \\
07360& 11.4& 16.7& 13.5& 1.6& $5.6\times10^7$& $6.0\times10^{-3}$& 6.9& 145 \\
\hline \hline
\end{tabular}
\end{center}
}
Explanation: The Table presents the results of the fit (cf: Fig~6)
of the jet surface brightness evolution to the function  
$(A_1 \delta (r) + A_2 r^{-\lambda} + A_3 (|-r|)^{-\lambda})$
convolved with 
$\frac{1}{\sqrt{2 \pi \sigma}} e^{- \frac{r^2}{2 \sigma ^2}},$
where $A_1$, $A_2$, and $A_3$ correspond to the core, jet, and counter
jet strengths, respectively, in arbitrary units; and $\sigma$ is fixed
according to the FWHM angular resolution. Values of $\chi^2$ are given 
corresponding to the power law fit, ${\chi_p}^2$, and the synchrotron loss
model, ${\chi_s}^2$.  
$T_b$ is the jet brightness temperature and $B_{eq}$ is the equipartition 
magnetic field in the jet both estimated  between 15 and 20 mas from the 
core, in order to avoid contamination by the core.
Since the jets are unresolved laterally, a jet width of 1 pc is assumed. 

\newpage

\begin{figure}[htb]
\vspace{7.0in}
\includegraphics{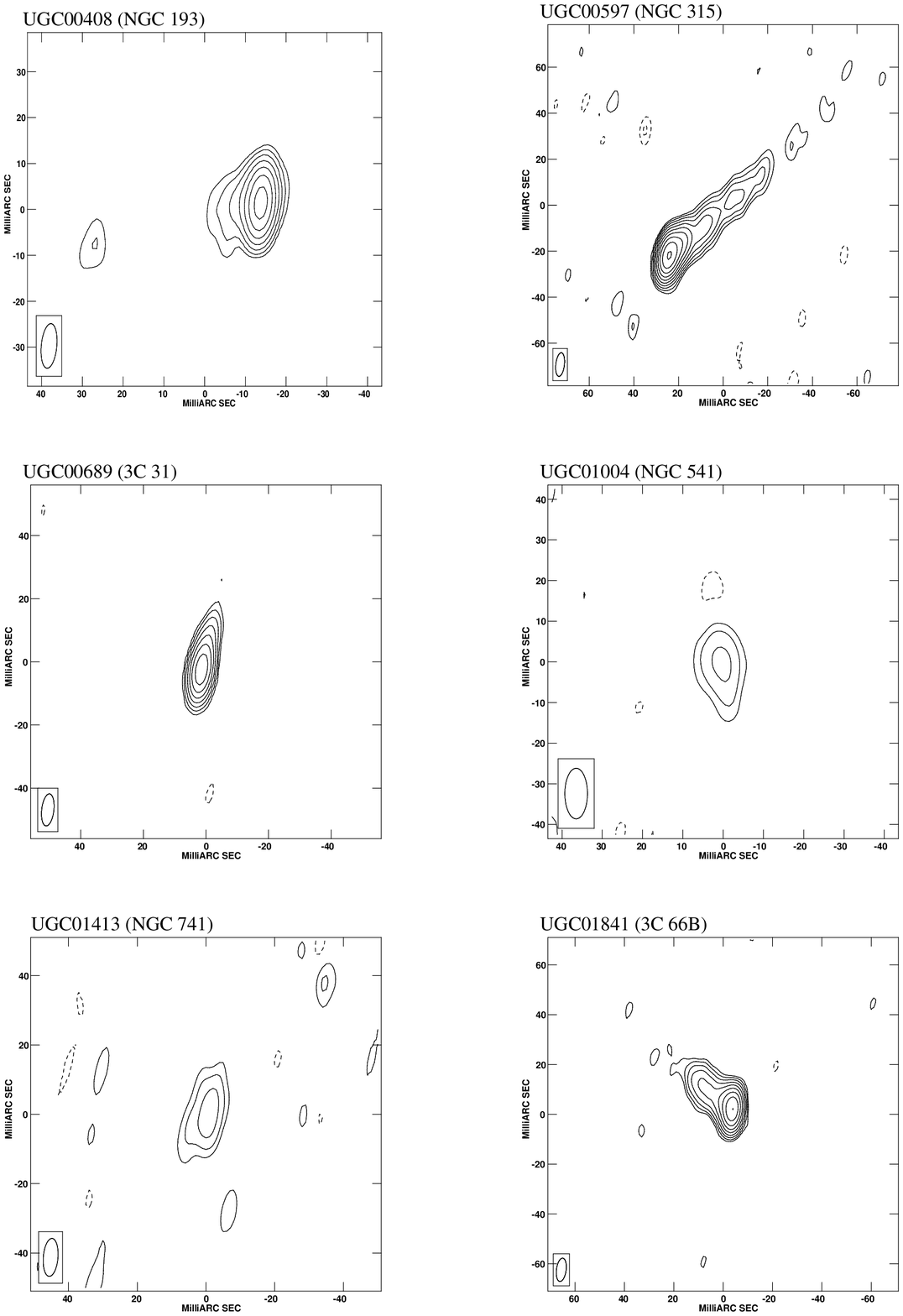}
\caption{}
\end{figure}

\begin{figure}[htb]
\vspace{7.0in}
\includegraphics{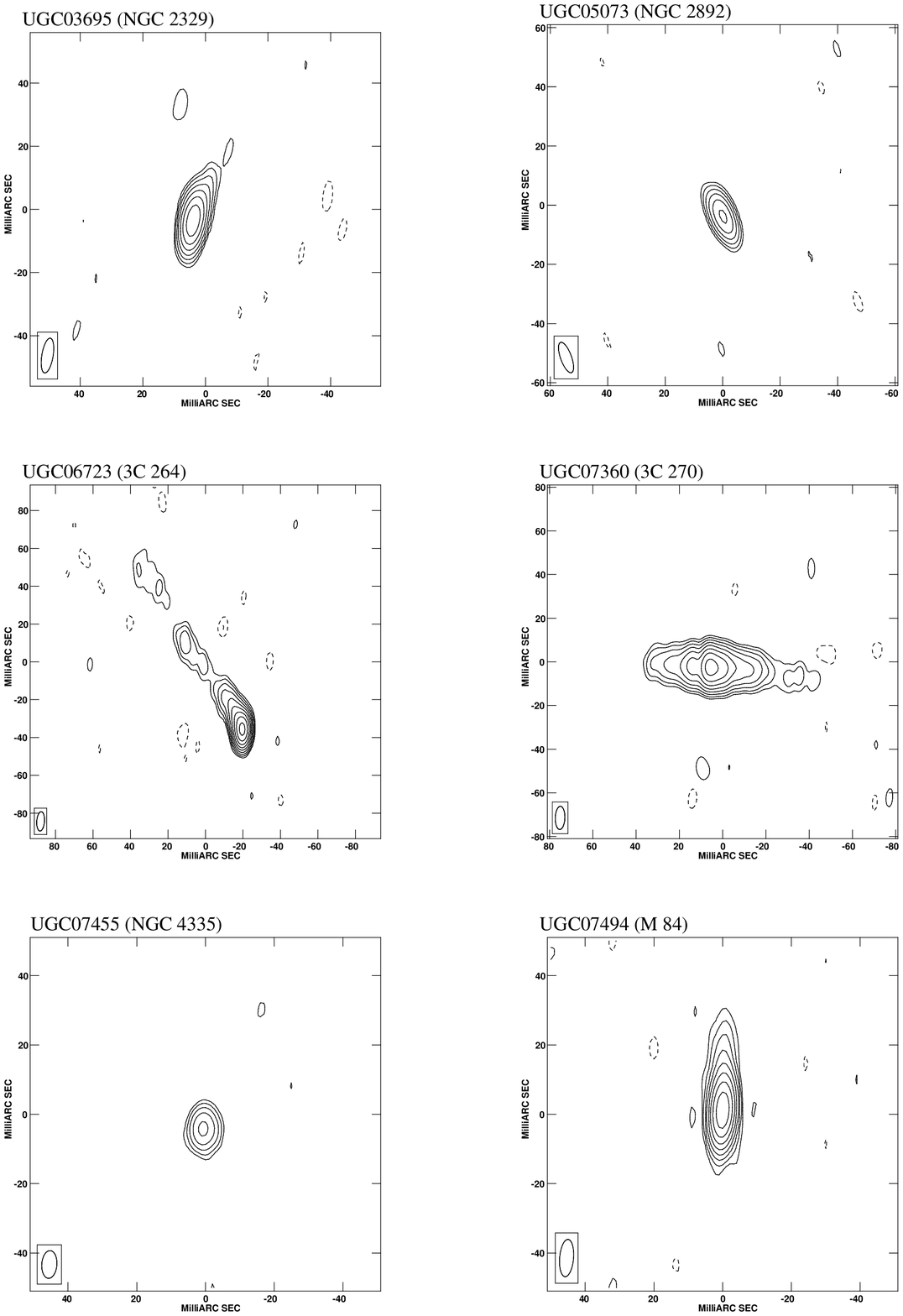}
\end{figure}

\begin{figure}[htb]
\vspace{7.0in}
\includegraphics{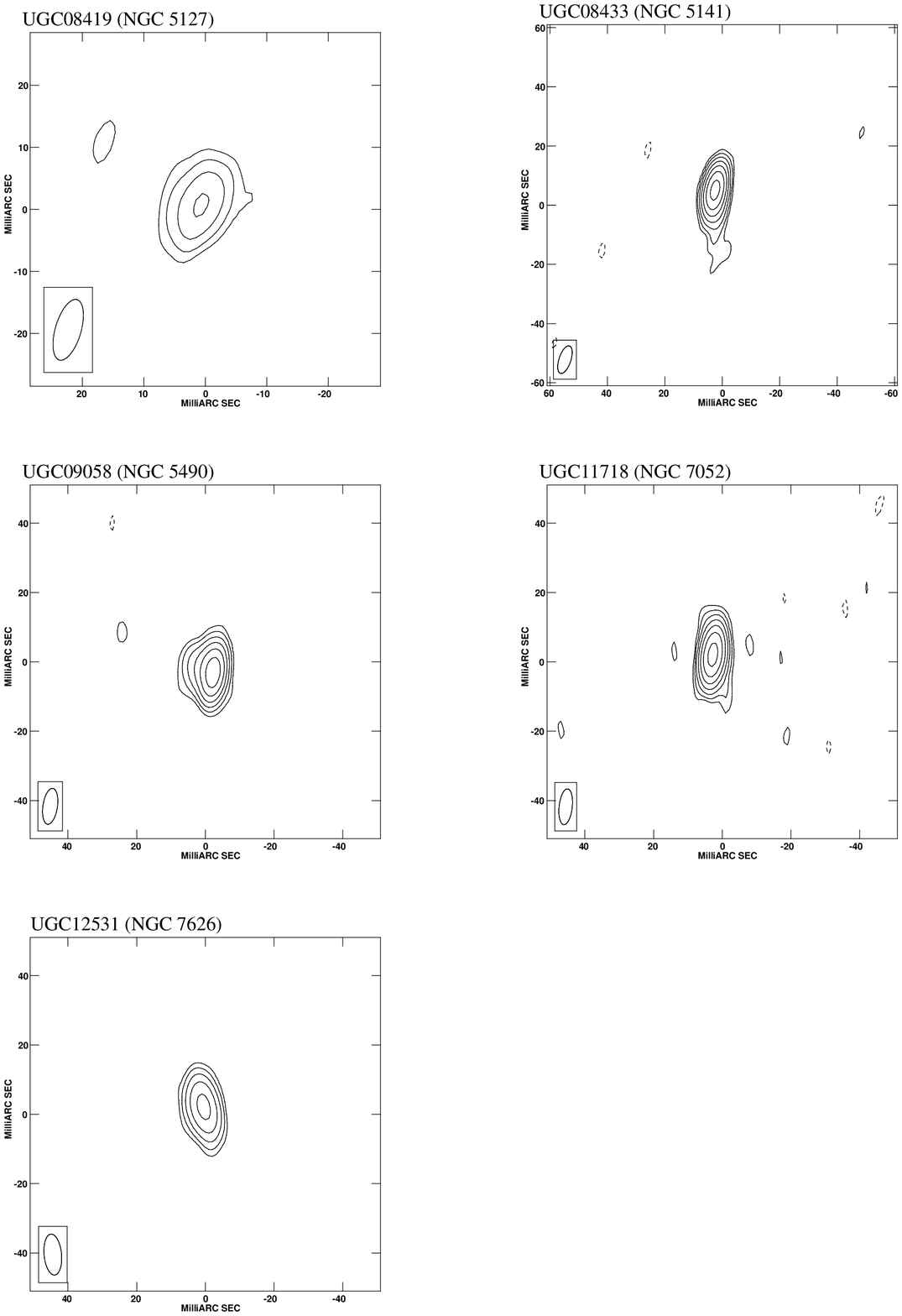}
\end{figure}

\begin{figure}[htb]
\vspace{7.0in}
\includegraphics{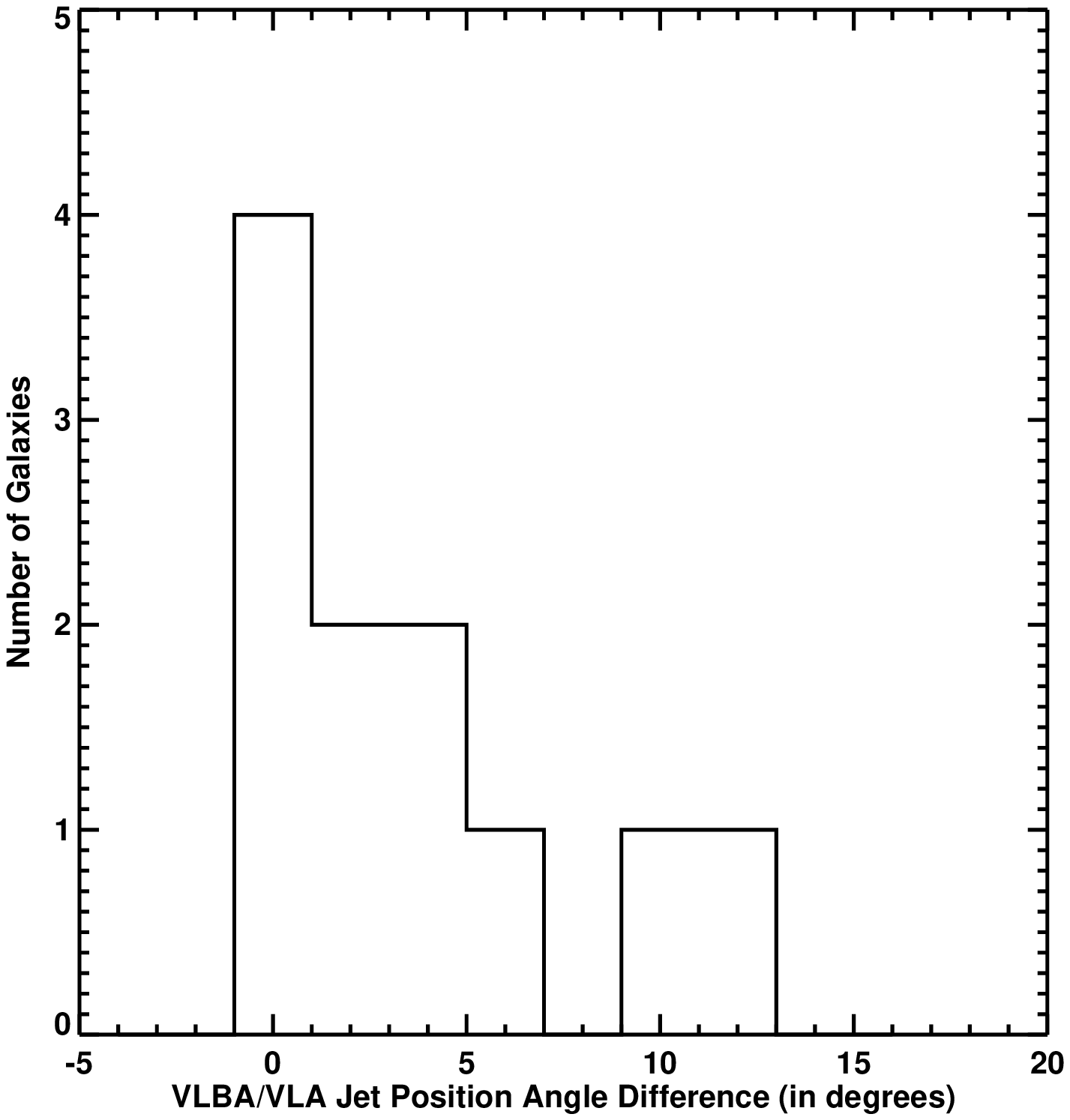}
\caption{}
\end{figure}

\begin{figure}[htb]
\vspace{7.0in}
\includegraphics{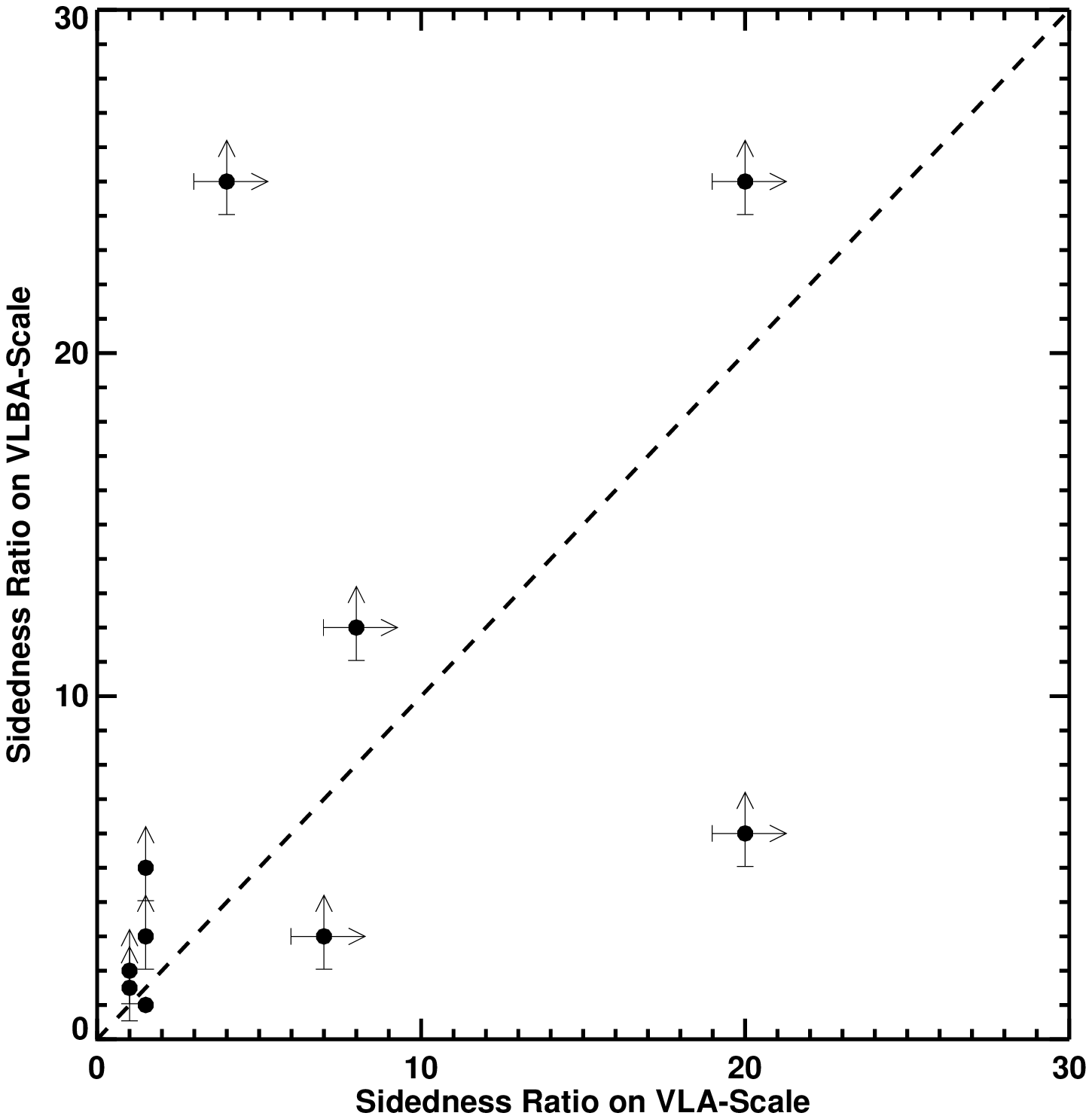}
\caption{}
\end{figure}

\begin{figure}[htb]
\vspace{7.0in}
\includegraphics{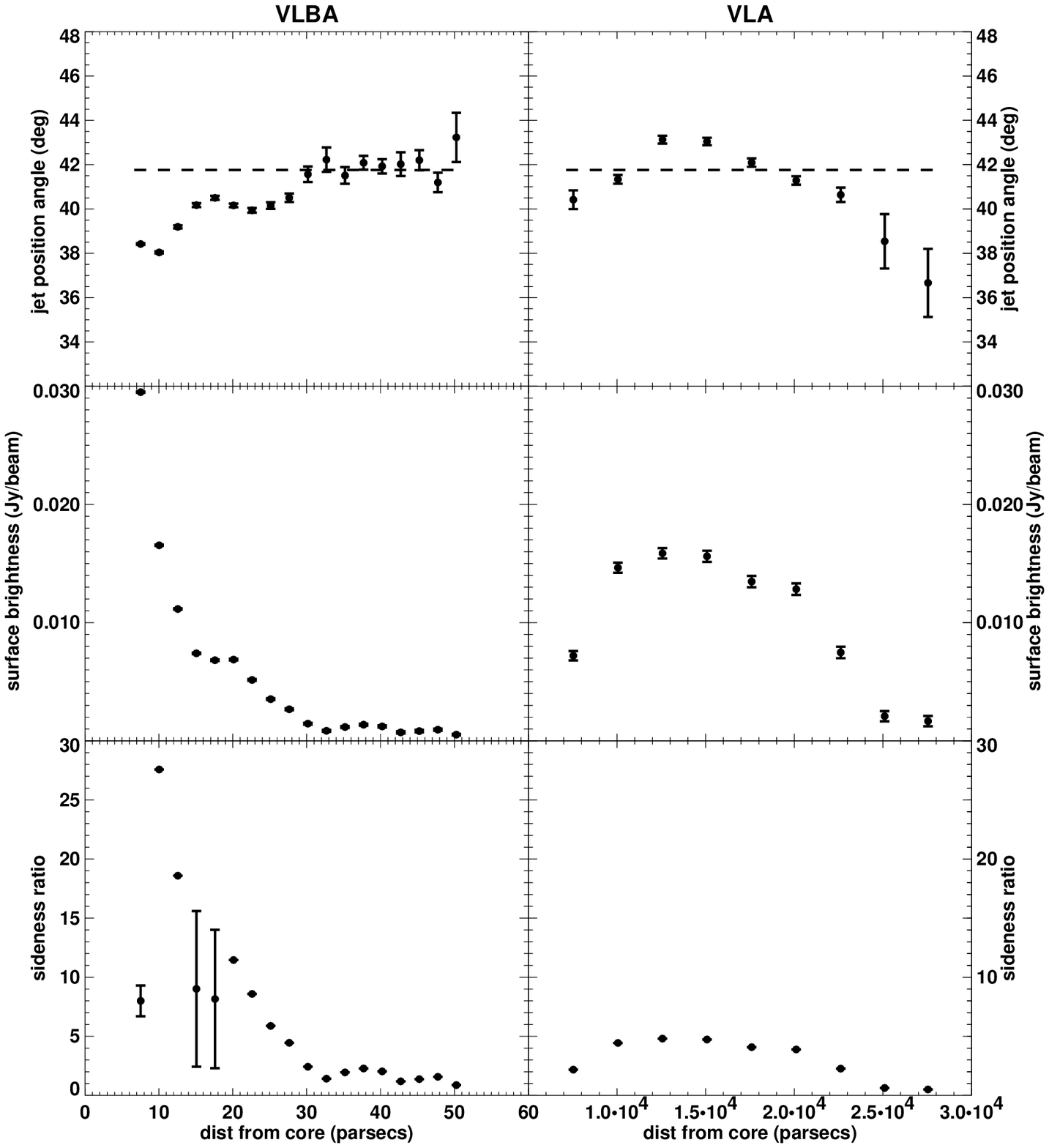}
\caption{}
\end{figure}

\begin{figure}[htb]
\vspace{7.0in}
\includegraphics{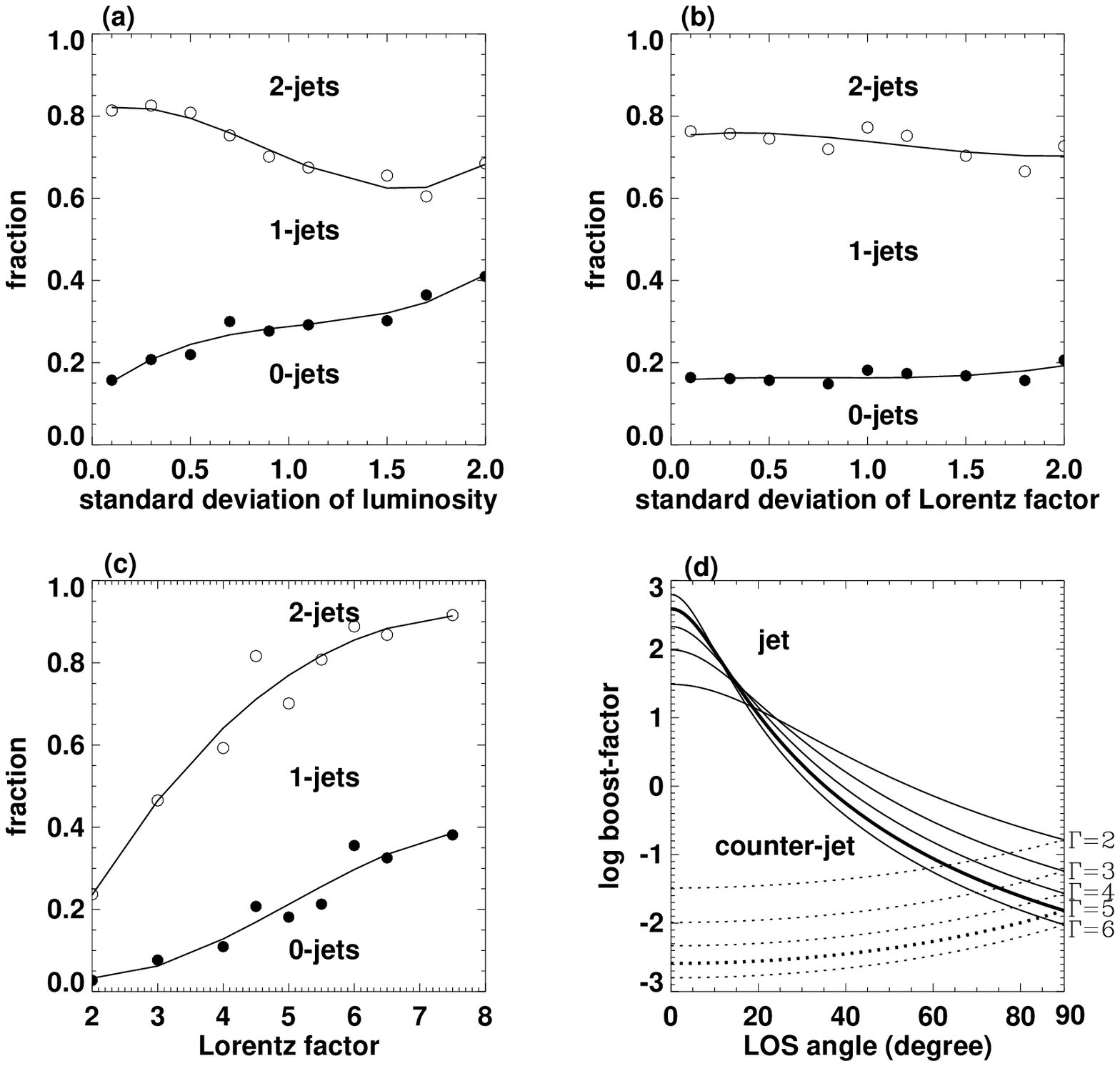}
\caption{}
\end{figure}

\begin{figure}[htb]
\vspace{7.0in}
\includegraphics{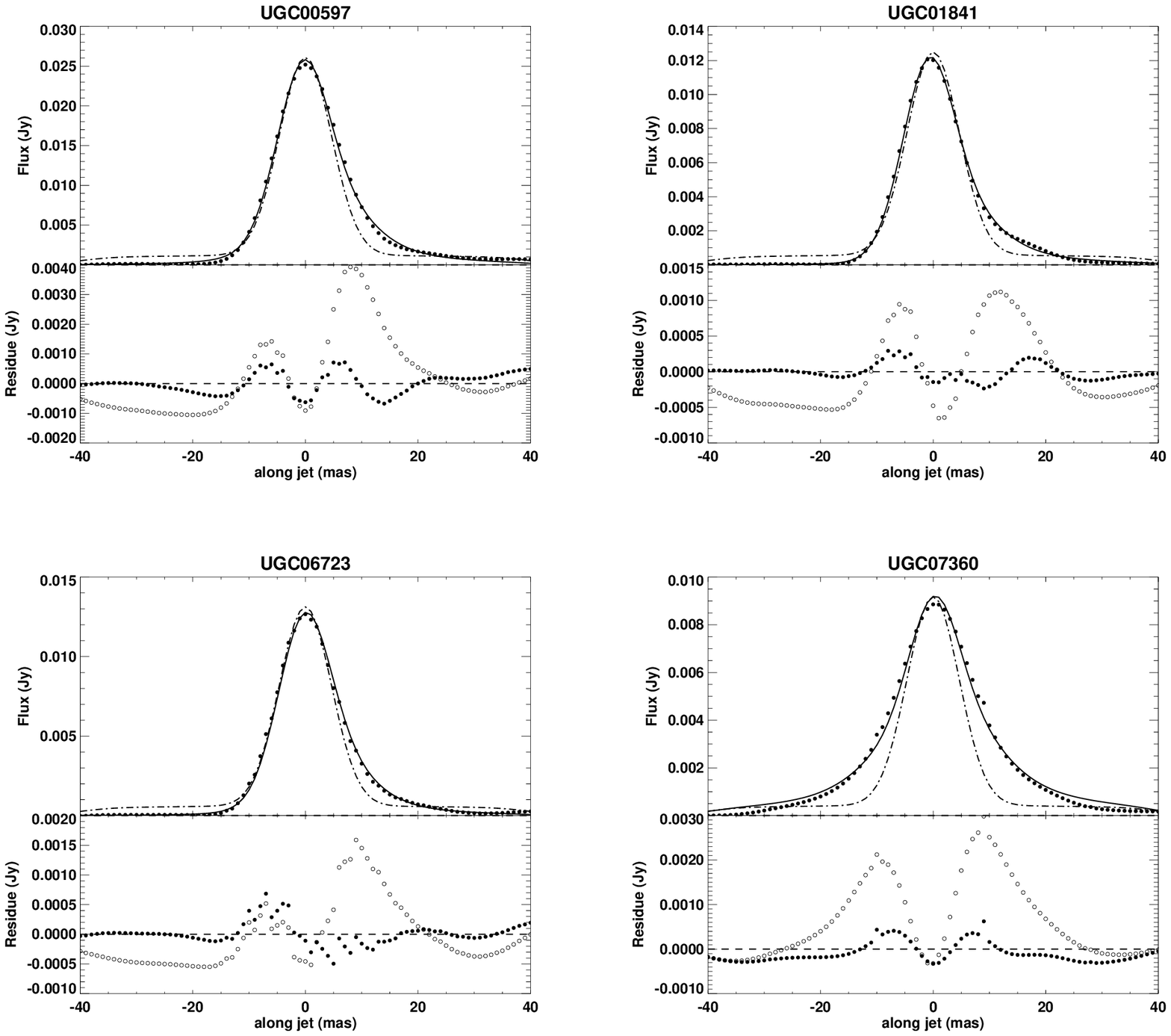}
\caption{}
\end{figure}

\begin{figure}[htb]
\vspace{7.0in}
\includegraphics{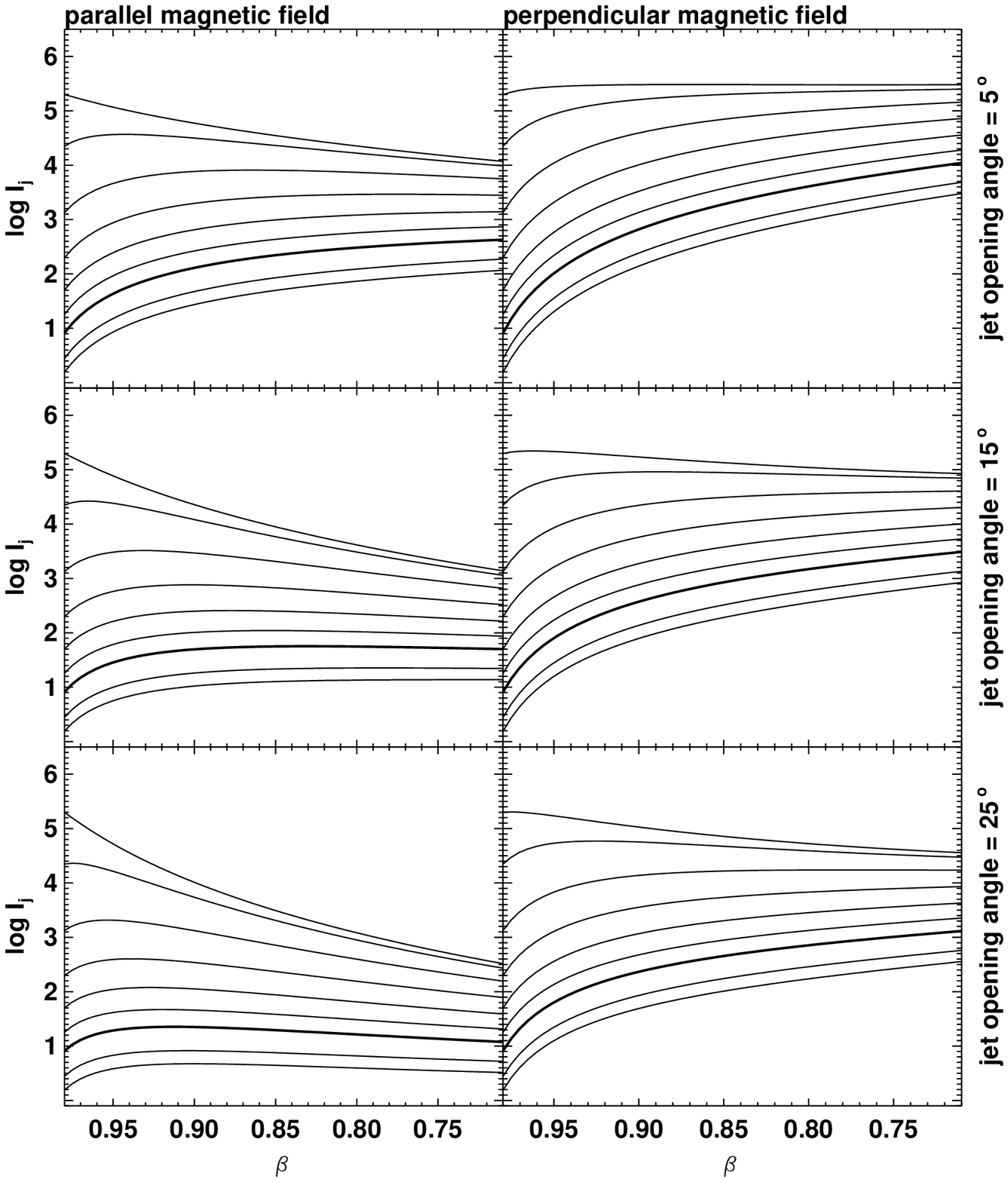}
\caption{}
\end{figure}

\end{document}